\documentclass[fleqn,usenatbib]{mnras}

\usepackage{newtxtext,newtxmath}

\usepackage[T1]{fontenc}
\usepackage{ae,aecompl}

\usepackage{graphicx}	
\usepackage{amsmath}	
\usepackage{amssymb}	
\usepackage{amsfonts}
\usepackage{epsfig,rotating}

\def\simeq{\mathrel{\raise.3ex\hbox{$\sim$}\mkern-14mu\lower0.4ex\hbox{$-$}}}

\def\ltsima{$\; \buildrel < \over \sim \;$}
\def\simlt{\lower.5ex\hbox{\ltsima}}
\def\gtsima{$\; \buildrel > \over \sim \;$}
\def\simgt{\lower.5ex\hbox{\gtsima}}

\def\msun{{\rm M_{\odot}}}

\def\be{\begin{equation}}
\def\ee{\end{equation}}
\def\ledd{{L_{\rm Edd}}}

\def\me{{\dot M_{\rm Edd}}}

\def\mbh{{M_{\rm BH}}}

\def\sgra{Sgr~A$^*$\,}
\def\del#1{{}}

\title[Sgr A* envelope explosion]{Sgr A$^*$ envelope explosion and the young stars in the centre of the Milky Way}

\author[S. Nayakshin, K. Zubovas]{Sergei Nayakshin$^{1}$\thanks{E-mail: sn85@le.ac.uk}
and Kastytis Zubovas$^{2,3}$
\\
$^{1}$Department of Physics and Astronomy, University of Leicester, University Road, LE1 7RH Leicester, United Kingdom\\
$^{2}$Center for Physical Sciences and Technology, Saul\.{e}tekio av. 3, Vilnius LT-10257, Lithuania \\
$^{3}$Vilnius University Observatory, Saul\.{e}tekio av. 3, Vilnius LT-10257, Lithuania
}

\date{Accepted XXX. Received YYY; in original form ZZZ}

\pubyear{2017}

\begin{document}
\label{firstpage}
\pagerange{\pageref{firstpage}--\pageref{lastpage}}
\maketitle

\begin{abstract}
\sgra is the super massive black hole residing in the centre of the Milky Way. There is plenty of observational evidence that a massive gas cloud fell into the central parsec of the Milky Way $\sim 6$~million years ago, triggering formation of a disc of young stars and activating \sgra.  In addition to the disc, however, there is an unexplained population of young stars on randomly oriented orbits. Here we hypothesize that these young stars were formed by fragmentation of a massive quasi-spherical gas shell driven out from \sgra potential well by  an energetic outflow. To account for the properties of the observed stars, the shell must be more massive than $10^5$ Solar masses,  be launched from inside $\sim 0.01$~pc, and the feedback outflow has to be highly super-Eddington albeit for a brief period of time, producing kinetic energy of at least $10^{55}$~erg. The young stars in the central parsec of the Galaxy may be a unique example of stars formed from atomic rather than molecular hydrogen, and forged by extreme pressure of black hole outflows.
\end{abstract}




\section{Introduction}

The central $\sim 2$~pc of the Milky Way is dominated in terms of mass by \sgra, the $\approx 4.3\times 10^6 \;\msun$ super-massive black hole \citep[SMBH; ][]{PaumardEtal06,GillessenEtal17}. Surprisingly, \sgra is orbited by over a hundred massive stars aged only $\sim 6$ million years \citep{LuEtal09}. This stellar population is confined strongly to the central $\sim 0.5$~pc \citep{YeldaEtal14}. Roughly 25\% of the young stars reside in a relatively well-defined disc \citep{Levin03} with stars on low eccentricity orbits \citep{BartkoEtal09,YeldaEtal14}. The disc of stars has an inner edge of $\sim 0.04$~pc and a top-heavy mass function with an unprecedented fraction of stellar mass in massive O/Wolf-Rayet stars \citep{NS05, PaumardEtal06}.

Self-gravitational collapse of a massive ($\sim 10^4 - 10^5 \msun$) gaseous disc explains the data for the stellar disc remarkably well. The disc cannot fragment inside $\sim 0.03$~pc \citep{NC05}, yields low eccentricity orbits \citep{AlexanderEtal08b}, and is expected to churn out very massive stars \citep{Levin03,Nayakshin06a}. 3D simulations \citep{BonnellRice08,HobbsNayakshin09} demonstrated how the gas disc forms via deposition of a massive gas cloud, and predicted that a fraction of the cloud would have low enough angular momentum to accrete onto \sgra. Strong \sgra activity $\sim 6$ million years ago is supported by the discovery of two $\sim 10$-kpc scale giant lobes emitting gamma-rays \citep[the {\em Fermi Bubbles};][]{SuEtal10}. Lobes of a similar shape and energy content form naturally as a result of feedback outflow launched by \sgra running into the ambient medium in the inner Galaxy \citep{ZubovasEtal11a,ZN12a}. The lobes were recently shown to be approximately coeval with the young stars \citep{MillerBregman16,BordoloiEtal17}, { disfavoring the competing star-formation feedback model for the origin of the Fermi lobes \citep{CrockerAharonian11,Crocker12}, which would make the lobes much older}.

In this paper we focus on the majority ($\sim 75\%$) non-disc population of young stars in the central parsec. Since they are on more isotropically distributed and more eccentric orbits, their formation cannot be explained by a gas disc fragmentation. Furthermore, inside the $\sim 0.04$~pc hole in the stellar disc,  there is at least a dozen of less massive B-type stars called S-stars \citep{Schoedel03,GhezEtal05}, which are even more eccentric (eccentricity $e\sim 0.4-0.9$), and are also isotropically distributed in the angular momentum directions \citep{GillessenEtal17}. For similar reasons, they too cannot be explained by a disc fragmentation \citep[ although some of the S-stars may migrate in from the larger scale stellar disc, see][]{Griv09}.
The leading scenario for formation of S-stars is tidal disruption of stellar binaries that pass too close to \sgra on nearly parabolic orbits \citep[e.g.,][]{Hills98}. However, \cite{HabibiEtal17} very recently found that the S-stars are co-eval with the disc stars within the errors,  challenging the binary disruption model. The model predicts post-disruption S-star eccentricities $e \sim 0.94 - 0.99$ \citep{Hills91,PeretsEtal09}. This is too large: relaxing these to the observed thermal eccentricity distribution \citep{GillessenEtal17} requires time at least an order of magnitude longer \citep[see Fig. 3 in][]{AlexanderT17} than the age of the S-stars. Additionally, in this model one also expects hundreds of B-type stars further out from \sgra, at distances $0.04 < R < 0.5$ pc, with very large eccentricities $\gtrsim 0.95$ \citep[see figs. 1-3 in][]{PG10}. The observed population is not as numerous or eccentric \citep{BartkoEtal09}.

Here we hypothesize that the non-disc population of young stars orbiting \sgra was formed via fragmentation of a very massive gas shell driven outward and compressed by \sgra feedback. Such shock-induced star formation is well known in the field of general star formation \citep{WhitworthEtal94,MachidaEtal05,ChiakiEtal13}, on scales of both individual star forming associations \citep{DehavrengEtal03,LiuT17} and also whole galaxies \citep{KetoEtal05}. Star formation inside AGN driven outflows was proposed recently by \cite{NZ12,Silk13}, and was possibly confirmed by \cite{MaiolinoEtal17} in a galactic outflow. 


\section{A model for S-star formation}

\subsection{Fragmenting shell} \label{sec:shell1}

Consider a spherical shell with thickness $Z$ much smaller than its radius $R$, with surface density $\Sigma$ and isothermal sound speed $c_s$. Approximating the shell as plane parallel, in the direction parallel to the shell, gravitational collapse of the fastest growing mode occurs on time scale \citep{WhitworthEtal94} $t_{\rm f} \sim 2 c_s/G\Sigma$,
and the corresponding linear size of the mode is
\begin{equation}
r_{\rm f} \sim \frac{2 c_s^2}{G\Sigma}\;.
\end{equation}
For the problem at hand, the shell is likely to be either expanding or contracting, depending on the balance of gravity, $F_{\rm g}/\left(4\pi R^2\right) = -G \mbh \Sigma/R^2$, and the outward pressure of \sgra outflow.
For self-gravitational collapse of the shell we therefore shall require that it occurred on a timescale similar to the local dynamical time, $t_{\rm dyn} = \Omega_K^{-1} = (R^3/G \mbh)^{1/2}$, i. e. 
\begin{equation}
Q_{\rm sh} \equiv \frac{t_{\rm f}}{t_{\rm dyn}} = \frac{2 c_s \Omega_K}{G \Sigma} \sim 1\;.
\end{equation}
Note that $Q_{\rm sh} = Q_{\rm T}/(2\pi)$, where $Q_{\rm T}$ is the \cite{Toomre64} parameter for a gas disc with same values of $\Sigma$ and $c_s$, so the collapse of the shell requires somewhat similar conditions albeit in a very different geometry. The mass of the shell fragment associated with the fastest growing mode is 
\begin{equation}
M_{\rm f} \sim \pi r_{\rm f}^2 \Sigma = 0.92\; \msun R_{0.01}^{3/2} \hat{T}^{3/2} Q_{\rm sh}\;,
\label{eq:mfrag0}
\end{equation}
where $\hat{T} = T/(3 \times 10^3$K) is the scaled gas temperature in the layer, and $R_{\rm 0.01} = R/(0.01$pc) is the distance from \sgra to the layer. This mass is much larger than the minimum mass of stars formed in ``normal'' Galactic conditions, $\sim 0.01 \msun$ \citep{Low76}.

The column density of the shell is 
\begin{equation}
\Sigma = \frac{2 c_s \Omega_K}{G Q_{\rm sh}} = 5.6\times 10^4 R_{0.01}^{-3/2} \hat{T}^{1/2} Q_{\rm sh}^{-1} \; {\rm g\ cm}^{-2} \;.
\end{equation}

While collapse starts initially in the plane of the shell, eventually it should turn into a 3D collapse. Therefore, we require the density of the shell, $\rho_{\rm sh}$, to at least exceed the tidal density,
\begin{equation} \label{eq:rhot}
\rho_{\rm sh} \gtrsim \rho_{\rm t} = \frac{\mbh}{2\pi R^3} \simeq 4.6 \times 10^{-11} R_{0.01}^{-3} {\rm g\ cm}^{-3}.
\end{equation}

The shell should also cool rapidly so that the compressional heat is removed before the internal pressure in the shell could resist collapse. The shell optical depth is $\tau = \kappa \Sigma$, where $\kappa(\rho_{\rm sh},T)$ is the Rosseland mean gas opacity at Solar metal abundance that we take from \cite{ZhuEtal09}. We shall assume that the shell temperature can be estimated via
\begin{equation}
T \sim T_{\rm bb} = \left(\frac{l L_{\rm Edd}}{4 \pi \sigma_{\rm SB} R^2}\right)^{1/4} \approx 5.2 \times 10^3 l^{1/4} R_{0.01}^{-1/2} K,
\end{equation}
where $T_{\rm bb}$ is the effective blackbody temperature at luminosity $l \ledd$ at distance $R$ away from \sgra, $\sigma_{\rm SB}$ is the Stefan-Boltzmann constant, and $l$ is a dimensionless parameter. The assumption $T\approx T_{\rm bb}$   is reasonable because at temperatures much higher than this the shell will cool down rapidly at distances commensurate with S-star orbits (see below). 

In the optically thin limit, gas clumps more massive than $\sim 0.01\msun$ can collapse dynamically due to rapid radiative cooling \citep{Rees76}. The collapse is considerably more difficult for an optically thick shell. We therefore consider this limit, when
the radiative cooling time of the shell is
\begin{equation}\label{eq:tcool}
t_{\rm cool} = \frac{\Sigma c_s^2 \tau}{\sigma_{\rm SB} T_{\rm bb}^4}\;.
\end{equation}

\begin{figure*}
\includegraphics[width=0.99\textwidth]{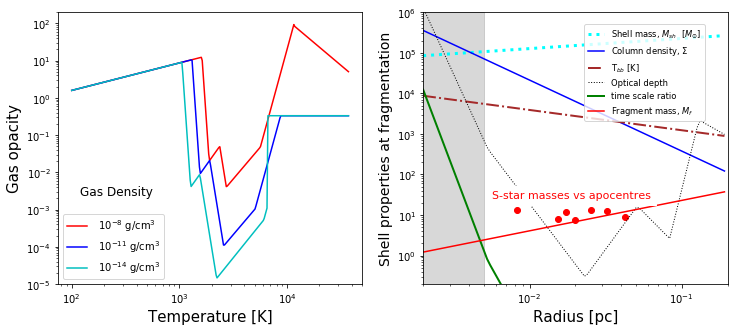}
\caption{{\bf Left:} Gas opacity $\kappa$ from Zhu et al (2009) as a function of gas temperature, for three different values of gas density. Note the opacity gap at temperature $2,000 \le T \le 10,000$~K. S-star formation could not occur without this opacity feature. {\bf Right:} Properties of the shell at collapse of the most unstable wavelength and the masses and apocentres of S-stars (filled circles; Gillessen et al 2017). S-star formation is only possible outside the shaded region, inside of which $t_{\rm cool}/t_{\rm dyn} > 1$.}
\label{fig:shell}
\end{figure*}

The left panel of Fig. \ref{fig:shell} shows the gas opacity as a function of temperature \citep[see][]{ZhuEtal09} for several values of gas density. The right panel of the figure shows various properties of the shell  calculated for $Q_{\rm sh}=3$. The green curve in particular shows the ratio $t_{\rm cool}/t_{\rm f}$. Only in the regions where the time scale ratio is less than unity could star formation take place. From this we conclude that the shell cannot collapse at radii smaller than $R_{\rm min} \sim 0.005$~pc (shaded region).  The transition between regions where star-formation is allowed and forbidden is very sharp because it corresponds to the $T\sim 10^4$ K ``wall'' of the opacity gap, 
where hydrogen atoms become ionized. Inside the gap the opacity is very low because hydrogen is almost all neutral or molecular, and the dust grains are also not present to contribute to the opacity.

The right panel in Fig. \ref{fig:shell} also shows the S-star mass versus apocentre of the orbits (red circles, see \S \ref{sec:obs} for why the apocentres are relevant here) for stars for which these quantities are both known \citep[see][]{GillessenEtal17,HabibiEtal17}. The red curve shows the fragment mass (eq. \ref{eq:mfrag0}). The agreement with observation in both S-star masses and allowed apocentres is surprisingly good given the approximate nature of our model, and should therefore be  considered somewhat fortuitous.

\subsection{Shell's origin: an explosion at Sgr A*?}\label{sec:origin}

The required mass of the shell, $M_{\rm sh} \gtrsim 10^5 \; \msun$, is comparable to the mass of the gas cloud invoked to be deposited in the central parsec \citep{BonnellRice08,HobbsNayakshin09} to explain the sub-parsec scale disc of young massive stars orbiting \sgra. To form the shell capable of making S-stars, however, we require this much gas to be deposited at  $R \sim 0.01$~pc. Nevertheless, this could occur if the {\em average} angular momentum of the cloud was small and significant angular momentum cancellation took place in self-collisional shocks \citep[e.g.,][]{HobbsEtal11}.

To estimate the rate of mass deposition into the $R \lesssim 0.01$~pc shell due to the cloud infall into the central parsec, we assume that gas fell towards \sgra at free-fall from $R_{\rm cl}\sim 0.5$~pc. This radius marks the outer edge of the stellar disc \citep{YeldaEtal14}, so we know there was a significant amount of gas deposited inside that region. We get
\begin{equation}
\dot{M}_{\rm dep} \sim \frac{M_{\rm cl}}{t_{\rm dyn}(R_{\rm cl})} \sim 40 M_5 \; R_{0.5}^{-3/2} \; \msun {\rm yr}^{-1} = 460\; \me\;,
\label{mdot0}
\end{equation}
where $M_5 = M_{\rm cl}/10^5\msun$, $R_{0.5} = R/(0.5$~pc), and $\me = 0.086 \;\msun$~yr$^{-1}$ is the Eddington accretion rate onto \sgra. 

\subsubsection{Accumulation from outside}\label{sec:outside-in}
 
 The infall rate in eq. \ref{mdot0} is very large, which raises the question: could the shell's significant mass accumulate in-situ, purely by infall from larger scales, while the shell is held up by feedback from \sgra a suitable distance from it, e.g., at $R \sim 0.01$~pc?

It is possible to arrange a feedback outflow from \sgra to produce just enough radial force to stop the material from falling into \sgra. However, $\sim 5000$~years are required to accumulate enough gas for shell fragmentation, whereas dynamical time at 0.01 pc is $\sim 7$~years. So the shell needs to be stable for a very long time and then become unstable for star formation. The feedback from \sgra would also need to increase with time to offset the increasing weight of the shell. This scenario is very finely tuned. Furthermore, the shell suspended some distance away from \sgra is unstable to fluid instabilities developing on short timescales, e.g., a few local dynamical times at the shell radius \citep{NZ12}. The shell would hatch dense filaments as the result of those instabilities. The filament weight per unit area is higher than the average for the shell, and they therefore fall deeper towards \sgra, despite the feedback emanating from the black hole. Since these instabilities develop on time scale of a few local dynamical times, before $\Sigma$ necessary for star formation in the shell is accumulated, we reject this scenario.

\subsubsection{Shell ejected from smaller scales}\label{sec:inside-out}

Let us consider the opposite possibility, that the shell came from much closer in, $R\ll 0.01$~pc. This could happen if gas falling from larger distances was deposited very close to \sgra in a massive accretion disc or a quasi-spherical envelope, and a fraction of that was then ejected due to an episode of super-Eddington activity of \sgra. 

\cite{BegelmanEtal08} studied ``quasi-stars'', massive quasi-spherical gas envelopes around stellar mass black holes in the very centres of young high-redshift galaxies. For these systems, the envelope mass greatly exceeds that of the black hole. In the case of \sgra, its gaseous envelope is unlikely to have been as massive as \sgra, at least not 6 million years ago,  but the physical principles governing the structure of the envelope are similar.

Quasi-stars  are strongly dominated by radiation pressure, and have outer radiative zones with temperature rising from a few $\times 10^3$~K on the envelope's outer radius, $R_{\rm e}$, to $\sim 10^5$~K at the convective-radiative boundary. The radiative zone can be large and its mass may be comparable to the total mass of the quasi-star \citep[eq. 29 in][]{BegelmanEtal08}. Since the luminosity of quasi-stars is limited by the Eddington luminosity, the outer radius of the envelope is given by
\begin{equation}
R_{\rm e} = \left[\frac{\ledd}{4\pi \sigma_{\rm SB} T_{\rm e}^4}\right]^{1/2} = 0.0075 \;{\rm pc }\; \left(\frac{\rm 6000 K}{T_{\rm e}}\right)^2
\label{Re0}
\end{equation}
where $T_{\rm e}$ is the envelope's effective temperature. The contraction (cooling) time of quasi-stars is very long compared with $t_{\rm dyn}$ at 0.5 pc.

3D radiative simulations of massive gas discs show that once magneto-rotational instability in the disc sets in, gas accretion rate onto the SMBH can rise much above the Eddington accretion rate \citep[e.g.,][finds accretion rates up to $\sim 1500\me$]{JiangEtal17}. The disc in this case becomes very geometrically thick, and an outflow is launched. \cite{JiangEtal17} also finds their discs strongly radiation-pressure dominated, and they find that magnetic field pressure also exceeds that of gas.

Such an unstable rapidly accreting inner disc with a very powerful outflow may form inside the quasi-star and eventually blow it apart from the inside. Since the quasi-star's optical depth is very large, radiation and energy released on small scales is trapped there, but the increased pressure will drive a nearly adiabatic expansion of the outer layers. Due to expansion, the temperature of these layers drops. Since the opacity of gas is such a strong function of temperature for $T \lesssim 10^4$~K, the outer layers of the star rapidly become optically thin once they cool below $10^4$~K, allowing radiation to leak out. This leads to a very large pressure drop in the outer layers, so that they can now be compressed to much higher density by the combined force of gravity and the outward acceleration from the expanding inner part of the quasi-star. This may then  lead to fragmentation, as described in Section \ref{sec:shell1}.

\subsection{Energetics and observational consequences}\label{sec:obs}

The proposed scenario for star formation results in strongly non-circular stellar orbits. In perfect spherical symmetry, stars born from the shell would be on exactly radial orbits, with eccentricity formally equal to 1 \citep[ we assume here that stars are born with velocities smaller than the escape velocity, although stars may be on escaping trajectories if the shell accelerates enough by the time it fragments, see][]{ZubovasEtal13a}. However, non-axisymmetric shell instabilities \citep{Vishniac1983,MacLowNorman93} result in additional, non-radial components to stellar velocities \citep[e.g., see figs. 1 \& 2 in][]{NZ12}, which would bring the eccentricity of newly made stars below 1, provided they remain bound to \sgra. Additional non-radial velocity components are expected if the quasi-star explosion itself is not spherical, e.g., if the shell surrounding \sgra is not perfectly spherical or if \sgra feedback is directed preferentially along SMBH's spin axis. Finally, magnetic fields might induce transverse motion in individual gas streams or clumps, reducing orbital eccentricity as the gas accumulates; these non-radial motions are destroyed by self-collision of gas streams, but some net angular momentum or turbulence might remain until the shell is blown away.

The semi-major axes of stars in this scenario depend on gas clump velocity at shell fragmentation. If clump velocity is significantly smaller than the local circular speed then the radius of the shell at fragmentation will set the apocentre of the stellar orbits. As the shell is driven outward, young stars may form on orbits with semi-major axes significantly larger than $0.01$~pc, perhaps accounting for all of the young stars in the inner half parsec of the Milky Way that are not in the stellar disc.

Pressure within the quasi-star bubble, $P_{\rm bub}$, needed to lift up the shell out of \sgra potential well is found from $4\pi R^2 P_{\rm bub} = (G \mbh/R^2) M_{\rm sh}$. Cast in units of ram pressure from an Eddington-limited momentum feedback outflow, 
\begin{equation}
\frac{4\pi R^2 P_{\rm bub}}{\ledd/c} =  \kappa_{\rm es} \Sigma = 2.2 \times 10^4 \; \hat{T}^{1/2} R_{0.01}^{-3/2} Q_{\rm sh}^{-1}\;,
\end{equation}
the pressure is very high. One can show that such a pressure is well above not only momentum-driven but also energy-driven {\em optically thin} AGN feedback outflows \citep[e.g.,][]{FEtal12a} limited by the Eddington luminosity, therefore requiring \sgra to be highly super-Eddington. 

The duration of the super-Eddington phase does not have to be long, however. The bubble minimum thermal energy is
\begin{equation}
E_{\rm bub} =  3 P_{\rm bub} V \sim 10^{55} \; {\rm erg} \;
\hat{T}^{1/2} R_{0.01}^{-1/2} Q_{\rm sh}^{-1}\;,
\label{E55}
\end{equation}
where $V = (4\pi/3) R^3$ is the bubble volume. This energy could be produced by accreting a rather modest amount of mass 
\begin{equation}
\Delta M = \frac{3 P_{\rm bub} V}{\varepsilon c^2} \approx 5000 \,
\msun \; \hat{T}^{1/2} R_{0.01}^{-1/2} Q_{\rm sh}^{-1} \epsilon_{-2}^{-1}
\end{equation}
where $V = (4\pi/3) R^3$ is the bubble volume and $\epsilon = 0.01 \epsilon_{-2}$ is feedback energy efficiency \citep[see][]{JiangEtal17}. For example, if \sgra accreted at 1000 times $\me$, then just $\sim 60$ years suffice to produce the needed energy. 
 In fact, a comparatively short duration of the quasi-star expansion phase is required for the self-consistency of the model. If the expansion takes many dynamical times at the outer edge of the quasi-star, pressure-deflated dense outer layers will become Rayleigh-Taylor unstable and fall through inside the quasi-star, just as was argued in \S \ref{sec:inside-out}. This is likely to drive very large scale convection on the outer edge rather than fragmentation. For this reason it is appropriate to call the hyper-Eddington expansion episode of the quasi-star an explosion.

 The velocity that the shell is ejected with is of the order of local escape velocity,
\begin{equation}
v_{\rm bub} \sim \left[ \frac{2G\mbh}{R}\right]^{1/2} = 1300 \;{\rm km\, s}^{-1}\; R_{0.01}^{-1/2}\;,
\end{equation}
although it can be larger if bubble expansion is accelerated by a continuous energy release by \sgra. The shell will also be slowed down when it runs into ambient interstellar medium.

These values for the outflow velocity and the kinetic energy are commensurable with the observational constraints from the Fermi Bubbles \citep{SuEtal10,ZubovasEtal11a}. The velocity kick that the gas in the Central Molecular Zone (CMZ) acquires due to interaction with the ejected shell is also of interest. Assuming that a fraction $\zeta \le 1$ of the shell's minimum momentum, $M_{\rm sh} v_{\rm bub}$, is passed on to the CMZ, which weighs $M_{\rm mz} \sim 5\times 10^7 \; \msun$, the kick is
\begin{equation}
v_{\rm k} \sim \zeta \frac{M_{\rm sh} v_{\rm bub}}{M_{\rm mz}} \sim 3\; \zeta\; {\rm km\, s}^{-1}\;,
\end{equation}
which is very much smaller than the circular velocity in the CMZ ($\sim 150$ km/s). This implies that the $\sim 200$~pc scale CMZ as a whole is not strongly affected by the shell ejection, although the smaller inner regions of the CMZ are much more susceptible to \sgra feedback.

\section{Discussion}\label{sec:obs}

We proposed that the quasi-spherical population of young stars in the central parsec of the Milky Way was formed inside a very dense shell of gas compressed and driven outward by a feedback outflow from \sgra. This mode of star formation is related to the AGN feedback induced star formation proposed recently \citep{NZ12,Silk13} and observed by \cite{MaiolinoEtal17} on much larger spatial scales. Note that the S-stars in this scenario are formed from gas dominated by atomic rather than molecular hydrogen, in contrast to the stars formed in the more benign conditions: even the clock-wise disc stars in the central parsec form out of molecular gas \citep{Levin03,Nayakshin06a}. S-star formation proposed here proceeds at much higher gas densities and temperatures as set by the properties of the expanding quasi-star.

The model predictions for S-star masses, semi-major axises and eccentricities are in a reasonable agreement with the observations (see \ref{fig:shell}).  The required shell mass is large, $M_{\rm sh} \gtrsim 10^5 \msun$, but perhaps could be lowered significantly if fragmentation occurred only in the filaments formed by the instabilities inside the shell \citep[see figs. 1 \&2 in][]{NZ12}. We also concluded that the shell must have originated from within the S-star orbits, requiring a very energetic, explosion-like feedback event very close to \sgra. 

Our hypothesis makes the S-stars and the quasi-spherical population of young stars further out physically distinct from the disc stars \citep{Levin03,PaumardEtal06}. However, we argue that these two populations are ultimately related through a {\em single} event that deposited a large quantity of gas into the central parsec \citep{BonnellRice08,HobbsNayakshin09,LucasEtal13}. The distinction would be the average angular momentum of the gas from which the two populations were made -- tiny for the quasi-spherical population, and finite, yielding the circularisation radius of $\sim 0.05-0.5$~pc, for the disc stars.

Numerical testing of this scenario for S-star formation requires 3D simulations that would model radiation-dominated fluid dynamics, and also resolve fluid instabilities to length scales much smaller than $0.01 R$. Stellar orbits predicted by such simulations can be then compared with the observed stellar orbits.
Another avenue to test our hypothesis is to model the ejected shell interaction with the well studied ambient gas distribution within the central Milky Way. 

We also expect similar black hole envelope explosions to occur in external galaxies. While these events may be very short lived compared to cosmological time scales, they should be longer in duration than the observable phases of supernovae, more luminous, and located at the very centres of the host galaxies.


\section*{Acknowledgements}

The authors thank Walter Dehnen, Andrew King, Reinhard Genzel, Stefan Gillessen and Maryam Habibi for discussions. KZ is funded by the Research Council Lithuania through the grant no. MIP-17-78. SN acknowledges support by STFC grant ST/K001000/1.




\bibliographystyle{mnras}
\bibliography{nayakshin}







\bsp	
\label{lastpage}
\end{document}